\documentclass[11pt,a4paper]{article}

\usepackage{amssymb}
\usepackage{amsmath}
\usepackage{subfigure}
\usepackage{graphicx}
\usepackage{fullpage}
\usepackage[footnotesize,bf]{caption}
\usepackage{mathptmx}

\newcommand{\eq}[1]{(\ref{#1})}
\newcommand{\bm}[1]{\boldsymbol{#1}}
\renewcommand{\d}{\mathrm{d}}
\newcommand{\Z}{\mathcal{Z}}

\begin{document}

\begin{center}
{\LARGE The $p(\gamma,K^+)\Lambda$ reaction:\\ consistent high-spin interactions and Bayesian inference of its resonance content}
\end{center}

%% use optional labels to link authors explicitly to addresses:
%% \author[label1,label2]{<author name>}
%% \address[label1]{<address>}
%% \address[label2]{<address>}

\begin{center}
T.~Vrancx\footnote{Electronic address: {\ttfamily tom.vrancx@ugent.be}}, L.~De~Cruz, J.~Ryckebusch, and P.~Vancraeyveld

\textit{\footnotesize Department of Physics and Astronomy, Ghent University, Proeftuinstraat 86, B-9000 Gent, Belgium}
\end{center}

\begin{abstract}
A Bayesian analysis of the world's $p(\gamma,K^+)\Lambda$ data is presented. We adopt a Regge-plus-resonance framework featuring consistent interactions for nucleon resonances up to spin $J = 5/2$. The power of the momentum dependence of the consistent interaction structure rises with the spin of the resonance. This leads to unphysical structures in the energy dependence of the computed cross sections when the short-distance physics is cut off with standard hadronic form factors. A plausible, spin-dependent modification of the hadronic form factor is proposed which suppresses the unphysical artifacts. Next, we evaluate all possible combinations of 11 candidate resonances. The best model is selected from the 2048 model variants by calculating the Bayesian evidence values against the world's $p(\gamma,K^+)\Lambda$ data. From the proposed selection of 11 resonances, we find that the following nucleon resonances have the highest probability of contributing to the reaction: $S_{11}(1535)$, $S_{11}(1650)$, $F_{15}(
1680)$, $P_{13}(1720)$, $D_{13}(1900)$, $P_{13}(1900)$, $P_{11}(1900)$, and $F_{15}(2000)$.
\end{abstract}

\setcounter{footnote}{0}	

\section{Introduction}
The Regge-plus-resonance (RPR) framework provides a hybrid model that conjoins an isobar description of nucleon exchanges in the $s$ channel and the exchange of Regge trajectories in the background channel. The members of a Regge trajectory share identical internal quantum numbers, such as strangeness and isospin, but have different total spins. More specifically, the spins and the squared masses of the Regge trajectory members are linearly related \cite{Corthals:2006}. As the $p(\gamma,K^+)\Lambda$ reaction is dominated by background contributions at forward kaon scattering angles, only $t$ channel Regge trajectories are introduced in the RPR model. The featuring Regge families of the RPR model are the $K^+(494)$ and $K^{*+}(892)$ trajectories. By replacing the Feynman propagators of the $K^+(494)$ and $K^{*+}(892)$ diagrams with the corresponding Regge propagators, the Reggeized background amplitude is obtained.

In the $s$ channel of the RPR model, individual nucleon resonances are exchanged. These nucleon resonances have half-integral spins and are modeled by effective Rarita-Schwinger (R-S) fields.\footnote{The spin-$1/2$ R-S field is the Dirac field. Only fields with a spin $J \ge 3/2$ will be referred to as R-S fields.} The finite lifetime of a resonance is incorporated in the RPR model by replacing the real pole of the Feynman propagator, i.e.\ $s-m_R^2$, with the imaginary pole $s - m_R^2 + i m_R\Gamma_R$. Here $m_R$ and $\Gamma_R$ represent the mass and the decay width of the resonance, respectively. In order to ensure that the resonance contributions vanish in the high-$s$ limit, a phenomenological hadronic form factor (HFF) is introduced. Such a form factor essentially ``cuts off'' the $s$ channel amplitude beyond a certain energy scale.

\section{Consistent interactions: RPR-2007 versus RPR-2011}
In Ref.\ \cite{Corthals:2007} a model was introduced that is now dubbed ``RPR-2007''. In this model, only spin-$1/2$ and spin-$3/2$ nucleon resonances are included. While spin-$1/2$ interactions are straightforward to deal with, spin-$3/2$ (in fact, spin $J \ge 3/2$ in general) couplings are more difficult to treat. The tensor-spinor representation of the R-S formalism inevitably invokes additional components for the R-S fields. These degrees of freedom are unphysical and are attributed to additional spin-$1/2$, $3/2$, $\ldots$, $J - 1$ components, next to the physical components of the spin-$J$ R-S field. It is clear that these unphysical degrees of freedom should be eliminated from the transition amplitude in order to have reaction observables that are physically meaningful.

\begin{figure}[t]
\centering
\includegraphics[width=.5\textwidth]{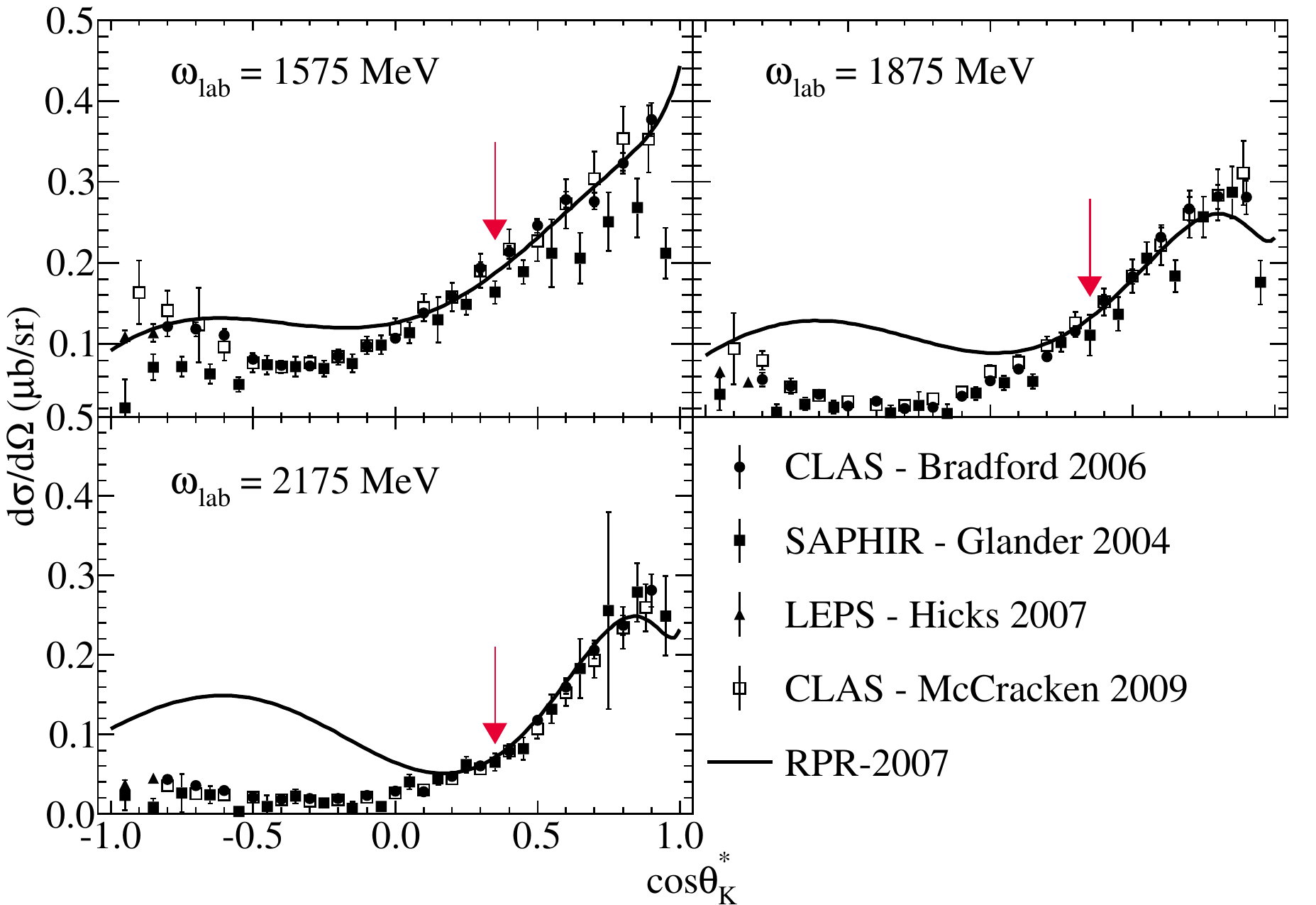}
\caption{RPR-2007 prediction and experimental data for the $p(\gamma,K^+)\Lambda$ differential cross section at photon lab energies of $1575$, $1875$, and $2175$ MeV. The RPR-2007 model is optimized against the $\cos\theta_K^* > 0.35$ data, which is indicated by the arrows. More details about the calculations and the data can be found in Ref.\ \cite{DeCruz:2012}.}
\label{RPR-2007_prediction}
\end{figure}

In RPR-2007 so-called ``contact-invariant'' interaction Lagrangians are used to characterize the spin-$3/2$ couplings \cite{Corthals:2007}. These Lagrangians contain one (two) ``off-shell'' parameter(s) for the strong (electromagnetic) vertex. There are hence three off-shell parameters for each amplitude that models a spin-$3/2$ resonance exchange. The HFF that is used in RPR-2007 is of the Gaussian form and has a common cutoff energy for all of the resonant amplitudes. The coupling constants, off-shell parameters, and the cutoff energy of the RPR-2007 model were optimized against the available, forward-angle experimental data at that time. The following resonance content was found for RPR-2007: $S_{11}(1650)$, $P_{11}(1710)$, $P_{13}(1720)$, $P_{13}(1900)$ and the missing resonance $D_{13}(1900)$. Fig.\ \ref{RPR-2007_prediction} shows the RPR-2007 prediction for the $p(\gamma,K^+)\Lambda$ differential cross section at three different photon energies in the lab frame. At forward kaon angles RPR-2007 is fully 
consistent with the data. At backward kaon angles, however, the model deviates significantly from the data. Moreover, an artificial bump is present and the situation worsens with increasing photon lab energy or, equivalently, increasing $s$. The observed elevation between the data and the RPR-2007 predictions can, to minor extent, be attributed to the fact that only $\cos\theta_K^* > 0.35$ data were considered when optimizing the RPR-2007 parameters. The main cause, however, of the unphysical bumps at $\cos\theta_K^* < 0$ is the adopted inconsistent interaction Lagrangians for the spin-$3/2$ vertices. As mentioned earlier, these Lagrangians contain one or two off-shell parameters. Now, these parameters allow for physical couplings to the unphysical components of the spin-$3/2$ R-S field. Hence, they give rise to unphysical structures in the predicted observables, which cannot be eliminated for any combination of the values of the off-shell parameters. What is more, the off-shell parameters are actually 
fitted to the data and, consequently, so are the artificial structures. Clearly this is a peculiar situation and it has to be put right.

In Ref.\ \cite{Vrancx:2011}, a formalism was developed in which the interaction of R-S fields with fundamental fields can be described in a consistent way. The consistent interaction Lagrangians are invariant under the so-called ``unconstrained R-S gauge'' and it was proven that the unphysical components of the R-S field decouple from the transition amplitude for this type of interactions. Moreover, the gauge-invariant Lagrangians do not contain any additional parameters, apart from the usual coupling constants. In Ref.\ \cite{Vrancx:2011} a novel, spin-dependent HFF was developed as well, dubbed the ``multidipole-Gauss'' HFF. This HFF is able to regularize the amplitudes involving consistently interacting R-S fields, something that is not feasible with the standard Gaussian HFF. The combination of consistent high-spin interactions and the accompanying multidipole-Gauss HFF, constitutes the foundations of a new RPR model, i.e.\ RPR-2011.

\begin{figure}
\centering
\subfigure[]{
\includegraphics[width=.4\textwidth]{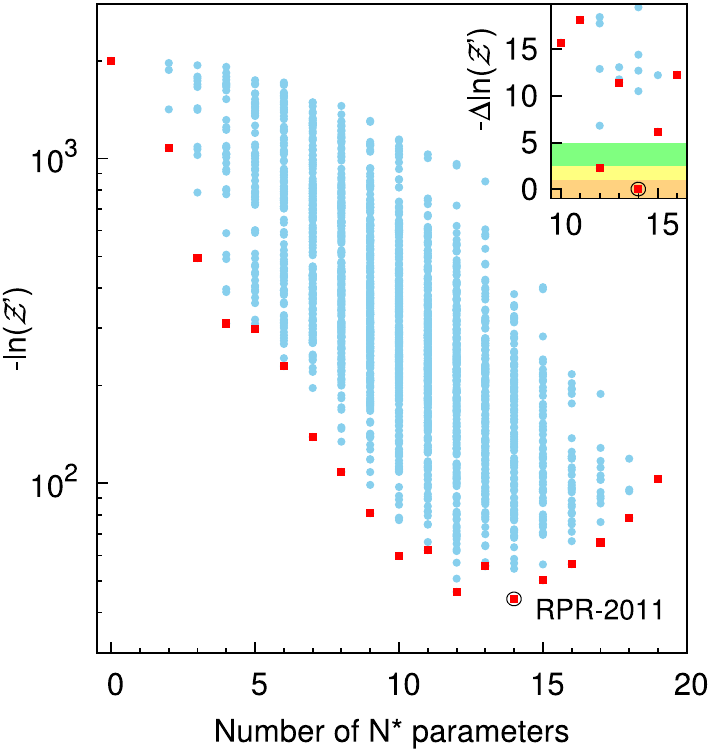}
}
\qquad
\subfigure[]{
\includegraphics[width=.5\textwidth]{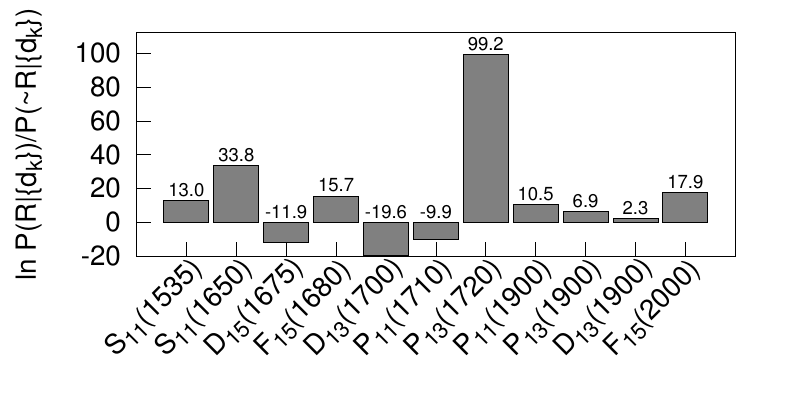}
}
\caption{(a) The evidence values for the 2048 RPR model variants (blue circles) in function of the number of nucleon resonance parameters. The most probable model for a fixed number of parameters is indicated with a red square. Top right inset: evidence ratios relative to RPR-2011 for the models with the highest corrected evidence. The color coding refers to Jeffreys' scale: barely worth mentioning (orange), significant
(yellow), strong to very strong (green) and decisive (white). More details can be found in Ref.\ \cite{DeCruz:2011} (b) The relative resonance probabilities for each of the 11 considered resonances. More details can be found in Ref.\ \cite{DeCruz:2012}.}
\label{best_rpr-estimate_ratio}
\end{figure}

\section{Bayesian inference of the RPR-2011 resonance content}
When it comes to specifying the set of nucleon resonances that has the most important contribution to the $p(\gamma,K^+)\Lambda$ reaction, one faces a lack of consensus between the various analyses for the reaction at hand. This disagreement becomes clear by inspecting Table I of Ref.\ \cite{DeCruz:2011}. The difficulty in determining the resonance contributions lies in the observed dominance of background (i.e.\ $t$ channel) reactions in the $p(\gamma,K^+)\Lambda$ process.

The challenge for the newly developed RPR model is to determine the most probable model variant $M$ (read: the most probable set of resonances) given the $p(\gamma,K^+)\Lambda$ data $\{d_k\}$ of the last decade. The standard $\chi^2$ distribution for the RPR model variant space is only capable of specifying the set of resonances that offers the best description of the data $\{d_k\}$. However, the $\chi^2$ distribution does not punish for the expansion of the set of resonances, i.e.\ the addition of extra model parameters. In fact, the absolute minimum of the $\chi^2$ hypersurface is most likely relocated, in general, by increasing the dimension of the model parameter space. Therefore, a model with a ``large'' number of contributing resonances is not a probable model.

Bayesian inference offers a quantitative way of selecting the most probable model amongst its possible variants. The Bayesian evidence $\Z$ for a specific set of resonances is defined as the probability of the data $\{d_k\}$, given the model variant $M$, and can be expressed as
\begin{align}
\Z = P(\{d_k\}|M) = \int P(\{d_k\},\bm{\alpha}_M|M) \d\bm{\alpha}_M, \label{Bayesian_evidence}
\end{align}
with $\bm{\alpha}_M$ being the model's parameters. The integrand of the right-hand side of Eq.\ \eq{Bayesian_evidence} is the product of the likelihood function $\mathcal{L}(\bm{\alpha}_M) = P(\{d_k\}|\bm{\alpha}_M,M)$ and the prior distribution $\pi(\bm{\alpha}_M) = P(\bm{\alpha}_M|M)$. The prior distribution for the model's parameters $\bm{\alpha}_M$ is chosen to be a uniform distribution. Now, consider the two model variants $M_A$ and $M_B$. The probabilities for both models are given by $P(\{ d_k \}| M_A)$ and $P(\{ d_k \}| M_B)$. By using Bayes' theorem, the probability ratio $P(\{ d_k \}| M_A)/P(\{ d_k \}| M_B)$ can be calculated as
\begin{align}
\frac{P(M_A|\{ d_k \})}{P(M_B|\{ d_k \})} = \frac{ P(\{ d_k \}| M_A)}{ P(\{ d_k \}| M_B)} \frac{P(M_A)}{P(M_B)} = \frac{ \mathcal{Z}_A}{\mathcal{Z}_B} \frac{P(M_A)}{P(M_B)}.
\end{align}
Since there is no prior preference for any of the two model variants, $P(M_A) = P(M_B)$ and the probability ratio reduces to the evidence ratio. The most probable model is therefore the model with the highest evidence. Due to the non-Gaussian and correlated nature of the systematic errors of $\{d_k\}$, the Bayesian evidence is underestimated. In Ref.\ \cite{DeCruz:2011} an approximate expression can be found for the corresponding corrected evidence, i.e.\ $\Z'$.

For the Bayesian analysis of the RPR-2011 model, 11 candidate resonances up to spin-$5/2$ are considered. So next to the consistent couplings, RPR-2011 differs from RPR-2007 in the inclusion of spin-$5/2$ resonances. The candidate resonances are: $S_{11}(1535)$, $S_{11}(1650)$, $D_{15}(1675)$, $F_{15}(1680)$, $D_{13}(1700)$, $P_{11}(1710)$, and $P_{13}(1720)$ (established resonances), $P_{13}(1900)$ and $F_{15}(2000)$ (less-estab\-lished resonances), and $D_{13}(1900)$ and $P_{11}(1900)$ (``missing'' resonances). Now, each of the $2^{11} = 2048$ model variants were optimized against $\{d_k\}$ (6148 data points to date) and the corresponding $\Z'$ value was calculated. The result is shown in Fig.\ \ref{best_rpr-estimate_ratio}(a).
The model variant with the highest evidence is indicated as ``RPR-2011''. The resonances that are not included in RPR-2011 are $D_{15}(1675)$,  
$D_{13}(1700)$, and $P_{11}(1710)$. The best RPR model variant, i.e.\ the one that includes all of the 11 candidate resonances, is hence not the most probable model. The individual probability ratio for each of the candidate resonances is shown in Fig.\ \ref{best_rpr-estimate_ratio}(b). A positive ratio indicates that it is more probable that the resonance contributes to the reaction $p(\gamma,K^+)\Lambda$ than that it does not. Vice versa, a negative ratio implies that the possibility that the resonance contributes to the reaction is not supported by the data. There are only 3 resonances that have a negative probability ratio, which are exactly those that are not part of the resonance set of RPR-2011.

\begin{figure}[t]
\centering
\includegraphics[width=\textwidth]{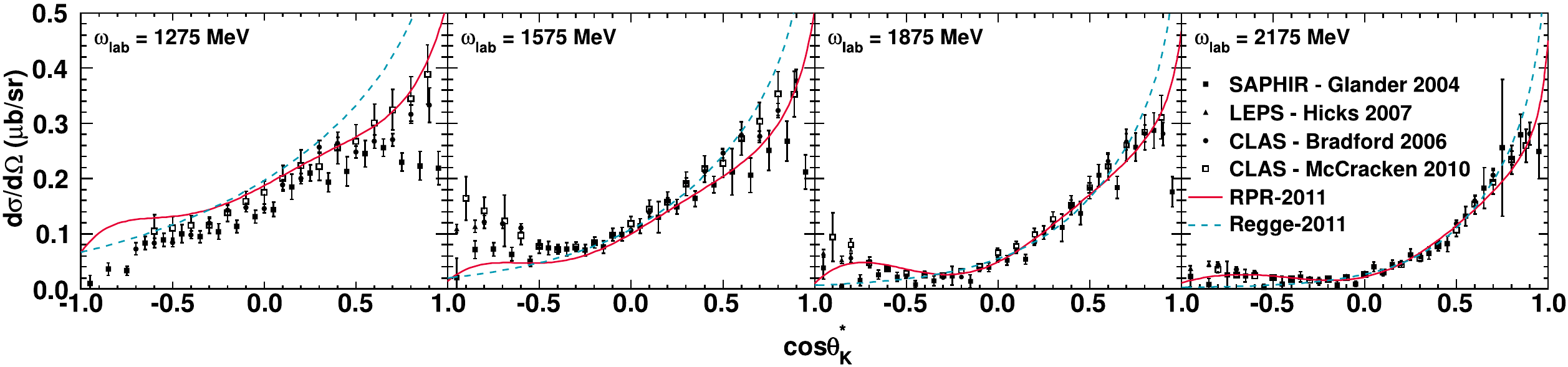}
\caption{RPR-2011 prediction and experimental data for the $p(\gamma,K^+)\Lambda$ differential cross section at photon lab energies of $1275$, $1575$, $1875$, and $2175$ MeV. More details about the calculations and the data can be found in Ref.\ \cite{DeCruz:2011}.}
\label{RPR-2011_prediction}
\end{figure}

In Fig.\ \ref{RPR-2011_prediction} the RPR-2011 prediction for the $p(\gamma,K^+)\Lambda$ differential cross section is shown at four different photon lab energies. It is seen that the bumps at $\cos\theta_K^* < 0$ (see Fig.\ \ref{RPR-2007_prediction}) are no longer present and that RPR-2011 offers a good description of the data for the whole range of kaon scattering angles.

\section{Acknowledgments}
This work is supported by the Research Council of Ghent University and the Flemish Research Foundation (FWO Vlaanderen).

\end{document}